\documentstyle[pmart,twoside,fleqn,epsf, amsmath, amssymb, graphicx]{article} % Do not change this line!

\begin{document}

\title{Reiter's Polaron Wavefunction  Applied \\ 
       to a $t_{2g}$ Orbital $t$--$J$ Model }
\author{Krzysztof Wohlfeld $^{a,b}$, Andrzej M. Ole\'s $^{a,b}$,
\\      Maria Daghofer $^{c} $, and Peter Horsch$^{b}$ }

\address{$^{a}$ Marian Smoluchowski Institute of Physics,
                Jagellonian University, \\
                Reymonta 4, PL--30059 Krak\'ow, Poland \\
         $^{b}$ Max-Planck-Institut FKF, Heisenbergstrasse 1,
                D--70569 Stuttgart, Germany \\
         $^{c}$ Department of Physics and Astronomy,
                The University of Tennessee, \\
                Knoxville, Tennessee 37996, USA }

%\date{May 5, 2008}

\maketitle         % Produces the title. This line is mandatory.


\begin{abstract}
Using the self-consistent Born approximation we calculate Reiter's
wavefunction for a single hole introduced into the undoped and
orbitally ordered ground state of the $t$--$J$ model with $t_{2g}$
orbital degrees of freedom. While the number of excitations is
similar to the spin $t$-$J$ model for a given $J/t$, a distinct
structure of the calculated wavefunction and its momentum
dependence is identified suggesting the formation of a novel type
of mobile polarons.
\end{abstract}

{\em Introduction---} Recently, using self-consistent Born
approximation (SCBA), we showed that a single hole introduced into
the undoped ground state of an orbital $t$--$J$ model with
$t_{2g}$ orbital degeneracy propagates coherently as a quasiparticle \cite{Dag08}. 
This striking result contradicts the na\"{i}ve expectations which suggest that a
hole should be trapped in this Ising-like ordered ground state. In
fact, the motion of a single hole is due to the frequently
neglected three-site terms and we showed \cite{Dag08} that this
new mechanism of hole movement is fundamentally different from the
coherent hole motion via quantum fluctuations in the standard spin $t$--$J$
model \cite{Mar91}, or in the case of $e_g$ orbitals \cite{vdB00}.
Though, a more detailed understanding of this novel mechanism is
needed. Hence, instead of considering the Green's function of the
problem \cite{Dag08}, we investigate the corresponding Reiter's
wavefunction \cite{Rei94} calculated in the SCBA.

{\em Orbital polarons---} The orbital $t$--$J$ Hamiltonian,
relevant for ferromagnetic $ab$ planes with two active $t_{2g}$
orbitals, $zx\equiv b $ and $yz\equiv a$, at each site, consists
of three terms \cite{Dag08} ${\mathcal H}={\mathcal H}_t+{\mathcal
H}_J+{\mathcal H}_{3s}$, with
\begin{subequations} \label{eq:H_tj2}
\begin{align}
\label{eq:Ht}
{\mathcal H}_{t}=&
-t \sum_{ i} \left( \tilde{b}^{\dagger}_{i}\tilde{b}^{}_{i+\bf{\hat{x}}} +
\tilde{a}^{\dagger}_{i}\tilde{a}^{}_{i+\bf{\hat{y}}} + h.c. \right), \\
\label{eq:HJ} {\mathcal H}_{J} =&\; \frac12 J \sum_{\langle
ij\rangle }
\left(T^z_i T^z_j - \frac{1}{4}\tilde{n}_i\tilde{n}_j\right),\\
\label{eq:H3s}
{\mathcal H}_{\rm 3s} =&
-\frac14 J \sum_{i} \left( \tilde{b}^\dag_{i-\bf{\hat{x}}}\tilde{n}^{}_{ia}
\tilde{b}^{}_{i+\bf{\hat{x}}} +
\tilde{a}^\dag_{i-\bf{\hat{y}}}\tilde{n}^{}_{ib}\tilde{a}^{}_{i+\bf{\hat{y}}}
+ h.c. \right),
\end{align}
\end{subequations}
where a tilde above fermion operators denote the restricted
Hilbert space without double occupancies, the total on-site
density
$\tilde{n}_i=\tilde{n}_{ia}+\tilde{n}_{ib}=\tilde{a}^\dag_i
\tilde{a}^{}_i+\tilde{b}^\dag_i \tilde{b}^{}_i$, and the
pseudospin operators $T^z_i=\frac{1}{2}
(\tilde{n}_{ib}-\tilde{n}_{ia})$. In Eq. (\ref{eq:H3s}) we
neglected the three-site terms which require orbital excitation
and therefore do not change the physical properties of the system
in the low doping regime (cf. discussion in Ref. \cite{Dag08}).

Following Ref. \cite{Mar91} we reduce the $t$--$J$ model to a
polaronic problem, which is a physically justified procedure for
the AF or AO ordered phases with low concentration of added holes
\cite{Bru00}. Hence, we: (i) divide the square lattice into two
sublattices $A$ and $B$, (ii) rotate pseudospins on the $A$
sublattice, and (iii) introduce fermion operators $h^\dag_i$ and
hard-core boson operators $\alpha^\dag_i$ such that
$\tilde{a}_i=h_i^\dag \alpha_i$ and
$\tilde{b}_i=h_i^\dag(1-\alpha_i^\dag\alpha_i)$. Then, in the
linear spin-wave approximation and having only one doped hole in
the plane, we obtain the following Fourier-transformed polaronic
Hamiltonian, $H= H_t+H_J+ H_{3s}$, with
\begin{subequations}
\label{Ht2g_pol}
\begin{align}
\label{Ht_pol} H_{t}\!=& \frac{1}{\sqrt{N}}\! \sum_{{\bf k},{\bf
q}} \! \left[ M_x({\bf k},\! {\bf q}) h^\dag_{{\bf k} A}h_{{\bf
k}-{\bf q} B}^{} \alpha_{{\bf q} A}^{}\!\! +\! M_y({\bf k},\!{\bf
q}) h^\dag_{{\bf k} B}h_{{\bf k}-{\bf q} A}^{} \alpha_{{\bf q}
B}^{}\! +\! h.c.  \right]\!,
\\
\label{HJ_pol} H_{J}\!=&\; \omega_0 \sum_{\bf k} \left(
\alpha_{{\bf k} A}^\dag \alpha^{}_{{\bf k} A} +
\alpha_{{\bf k} B}^\dag \alpha_{{\bf k} B}^{} \right),\\
\label{H3s_pol} H_{\rm 3s}\!=&\; \sum_{\bf k} \left[ \varepsilon_y
({\bf k}) h_{{\bf k} A}^\dag h_{{\bf k} A}^{} \!+\! \varepsilon_x
({\bf k}) h_{{\bf k} B}^\dag h_{{\bf k} B}^{} \right].
\end{align}
\end{subequations}
Here the sums go over all momenta $\{{\bf k}\}$ in the Brillouin
zone for the whole lattice\footnote{One can also perform the sum
only in the reduced Brillouin Zone although this requires to
change $N \to N/2$ everywhere before summations.}, the total
number of sites is $N$, and the orbiton energy is $\omega_0 = J$.

The vertices and the dispersion relations are equal to (with
$\nu= x,y$):
\begin{equation} \label{eq:disp}
M_{\nu}({\bf k}, {\bf q}) = 2 t \cos (k_{\nu} -q_{\nu}),
\hskip 1.7cm \varepsilon_{\nu} ({\bf k}) =
\textstyle{\frac{1}{2}} J \cos (2k_{\nu}).
\end{equation}

{\em Reiter's wavefunction---} In the spirit of Ref. \cite{Rei94},
let us assume that the wavefunction for a hole with quasiparticle
(QP) momentum ${\bf k}$ and initially doped into the sublattice
$A$ and $B$, respectively, takes the form:
\begin{subequations}
\begin{align}
\label{eq:ansatzA} |\Psi_{{\bf k}A}\rangle=&\; a^{(0)}_A({\bf
k})h^\dag_{{\bf k}A}|0\rangle + \frac{1}{\sqrt{N}} \sum_{\bf q}
a^{(1)}_A({\bf k},{\bf q})h^\dag_{{\bf k}-{\bf q}B}
\alpha_{{\bf q} A}^\dag|0\rangle \nonumber \\
+&\;\frac{1}{N} \sum_{{\bf q}_1,{\bf q}_2} a^{(2)}_A({\bf k},{\bf
q}_1,{\bf q}_2) h^\dag_{{\bf k}-{\bf q}_1-{\bf q}_2 A}
\alpha_{{\bf q}_1A}^\dag \alpha_{{\bf q}_2B}^\dag
|0\rangle+\cdots\ ,
\end{align}
\begin{align}
\label{eq:ansatzB}
|\Psi_{{\bf k}B}\rangle=&\; a^{(0)}_B({\bf
k})h^\dag_{{\bf k}B}|0\rangle + \frac{1}{\sqrt{N}} \sum_{\bf q}
a^{(1)}_B({\bf k},{\bf q})h^\dag_{{\bf k}-{\bf q}A}
\alpha_{{\bf q}B}^\dag|0\rangle \nonumber \\
+&\;\frac{1}{N} \sum_{{\bf q}_1,{\bf q}_2} a^{(2)}_B({\bf k},{ \bf
q}_1,{\bf q}_2) h^\dag_{{\bf k}-{\bf q}_1-{\bf q}_2B} \alpha_{{\bf
q}_1B}^\dag \alpha_{{\bf q}_2A}^\dag |0\rangle+\cdots \ .
\end{align}
\end{subequations}
The coefficients $\Big|a^{(0)}_L({\bf k})\Big|^2$ (with the sublattice index
$L\in \{A,B\}$) are the QP spectral weights which follow from the
normalization of the wavefunction, whereas $a^{(n)}_L({\bf k},{\bf
q}_1,..., {\bf q}_n)$ for $n>0$ are to be determined from the
Schr\"odinger equations $H |\Psi_{{\bf k}L}\rangle=\lambda_{{\bf
k}L} |\Psi_{{\bf k}L}\rangle$. Substituting Eq.
(\ref{eq:ansatzA})--(\ref{eq:ansatzB}) into them yields
\begin{subequations}
\label{eq:coeff}
\begin{eqnarray}
\label{eq:coeff1}
a^{(2n)}_A\!({\bf k}\!,\!\{{\bf q}_i\}_{ 2n})\!\!\! &\!\!\! =\!\!\!&\!\!\!
a^{(2n-1)}_A\!({\bf k},\!\{{\bf q}_i\}_{2n\!-\!1})\!M_y\!(\bar{{\bf k}}_
{2n\!-\!1}\!,\!{\bf q}_{2n})G_{A}(\bar{{\bf k}}_{2n},\!\mu^{(2n)}_{{\bf k}A}),
\\
\label{eq:coeff2}
 a^{(2n-1)}_A({\bf k},\!\{{\bf q}_i\}_{2n\!-\!1})\!\!\! &\!\!\! =\!\!\!&\!\!\!
a ^{(2n\!-\!2)}_A\!({\bf k},\!\!\{{\bf q}_i\}_{2n\!-\!2}\!)
M_x(\bar{{\bf k}}_{2n\!-\!2},{\bf q}_{2n\!-\!1})G_{B}(\bar{{\bf k}}_{2n\!-\!1},
\mu^{(2n\!-\!1)}_{{\bf k}A}),
\nonumber \\
\\
\label{eq:coeff3}
a^{(2n)}_B\!({\bf k}\!,\!\{{\bf q}_i\}_{ 2n})\!\!\! &\!\!\! =\!\!\!&\!\!\!
a^{(2n-1)}_B\!({\bf k},\!\{{\bf q}_i\}_{2n\!-\!1})
\!M_x\!(\bar{{\bf k}}_{2n\!-\!1}\!,\!{\bf q}_{2n})G_{B}(\bar{{\bf k}}_{2n},
\!\mu^{(2n)}_{{\bf k}B}),
\\
\label{eq:coeff4}
 a^{(2n-1)}_B({\bf k},\!\{{\bf q}_i\}_{2n\!-\!1}) \!\!\! &\!\!\! =\!\!\!&\!\!\!
a ^{(2n\!-\!2)}_B\!({\bf k},\{{\bf q}_i\}_{2n\!-\!2} )
M_y(\bar{{\bf k}}_{2n\!-\!2},{\bf q}_{2n\!-\!1})G_{A}(\bar{{\bf k}}_{2n\!-\!1},
\mu^{(2n\!-\!1)}_{{\bf k}B}),\nonumber \\
\end{eqnarray}
\end{subequations}
where $\bar{\bf k}_n={\bf k}-{\bf q}_1-...-{\bf q}_n$, $\{{\bf
q}_i\}_{n}=\{{\bf q}_1, {\bf q}_2, ..., {\bf q}_{n} \}$,
$\mu^{(n)}_{{\bf k}L} = \lambda_{{\bf k}L} - n \omega_0$. The
Green's functions $G_L$ are defined by the self-consistent
equations
\begin{subequations}
\label{eq:green}
\begin{align}
G^{-1}_A({\bf k}, \omega)&=\omega-\varepsilon_y({\bf k})-\frac{1}{N}
\sum_{\bf q} G_B({\bf k}-{\bf q}, \omega- \omega_0) M_x^2 ({\bf k},{\bf q})\ , \\
G^{-1}_B({\bf k}, \omega)&=\omega-\varepsilon_x({\bf
k})-\frac{1}{N} \sum_{\bf q} G_A({\bf k}-{\bf q}, \omega-
\omega_0) M_y^2 ({\bf k},{\bf q})\ ,
\end{align}
\end{subequations}
and the QP energy has to be equal
\begin{subequations}
\label{eq:qpenergy}
\begin{eqnarray}
\lambda_{{\bf k}A} &=& \varepsilon_y({\bf k}) + \frac{1}{N}
\sum_{\bf q} G_B ({\bf k}-{\bf q},
\mu^{(1)}_{{\bf k}A}) M_x^2({\bf k},{\bf q})\ , \\
\lambda_{{\bf k}B} &=& \varepsilon_x({\bf k}) + \frac{1}{N}
\sum_{\bf q} G_A ({\bf k}-{\bf q}, \mu^{(1)}_{{\bf k}B})
M_y^2({\bf k},{\bf q})\ .
\end{eqnarray}
\end{subequations}

Note, that in order to obtain equations for the coefficients of
the Reiter's wavefunction $a^{(n)}_L$  Eqs. (\ref{eq:coeff}) we
adopted the following contraction procedure: we neglected all
terms which would correspond to the annihilation of orbitons in
the $n$-th step if the orbitons were created earlier than in the
$(n-1)$-th step. This procedure resembles the non-crossing
approximation while calculating Green's function using the
diagrammatic technique. In fact, one may wonder why we still have
to adopt such approximation since the closed loops which
correspond to the crossing diagrams are anyway prohibited in the
orbital $t$--$J$ model under consideration \cite{Dag08}. The
answer to this puzzle is the following: to conclude that the
crossing diagrams are unphysical we not only need to look at the
structure of the Hamiltonian Eq. (\ref{Ht2g_pol}) but also at the
processes allowed by this Hamiltonian on the square lattice. Thus,
we need some extra information about the lattice to exclude the
crossing diagrams, and consequently we need to introduce the
contraction procedure by hand.

{\em Consequences---} Eqs. (\ref{eq:coeff}) together with Eqs.
(\ref{eq:green}) and (\ref{eq:qpenergy}) can be easily solved
numerically. While it is impossible to calculate the coefficients
$a^{(n)}_L$ for all $n$, one may ask whether there exists such an
$m$ that all the coefficients with $n>m$ are so small that they
could be safely neglected. This question is also of physical
importance since knowing $m$ would mean that the wavefunction of
the doped hole can be approximated by a superposition of the
wavefunction of a free hole and $m$ wavefunctions of the hole
dressed with $m$ orbitons.

The easiest way to answer the above question is to calculate the
norm
\begin{align}
&N_{{\bf k}A}\equiv \langle \Psi_{{\bf k}A}|\Psi_{{\bf k}A}\rangle
= \Big|a^{(0)}_A({\bf k})\Big|^2 \Big\{ 1+\frac{1}{N}\sum_{{\bf q}_1} |
M_x({\bf k},{\bf q}_1)
G_{B}(\bar{\bf k}_1,\mu^{(1)}_{{\bf k}A})|^2 \nonumber \\
&+\frac{1}{N^2}\sum_{{\bf q}_1, {\bf q}_2} | M_x({\bf k},{\bf
q}_1) G_{B}(\bar{{\bf k}}_1,\mu^{(1)}_{{\bf k}A})|^2|
M_y(\bar{{\bf k}}_1,{\bf q}_2)
G_{A}(\bar{{\bf k}}_2,\mu^{(2)}_{{\bf k}A})|^2 \nonumber \\
&+\frac{1}{N^3}\sum_{{\bf q}_1, {\bf q}_2, {\bf q}_3} | M_x({\bf
k},{\bf q}_1) G_{B}(\bar{{\bf k}}_1,\mu^{(1)}_{{\bf k}A})|^2
|M_y(\bar{{\bf k}_1},{\bf q}_2)G_{A}(\bar{{\bf
k}}_2,\mu^{(2)}_{{\bf k}A})|^2
\nonumber \\
&\hskip .3cm | M_x(\bar{{\bf k}}_2,{\bf q}_3)G_{B}(\bar{{\bf
k}}_3,\mu^{(3)}_{{\bf k}A})|^2+\cdots \Big\} \equiv  N_{{\bf
k}A}^{(0)}\!+\!N_{{\bf k}A}^{(1)}\!+\!N_{{\bf
k}A}^{(2)}\!+\!N_{{\bf k}A}^{(3)}\!+\cdots\ ,
\end{align}
of the wavefunction on the $A$ sublattice given by Eq.
(\ref{eq:ansatzA}), and similarly $N_{{\bf k}B}$ for the $B$
sublattice. Naturally, $N_{{\bf k}L} = 1$ which yields the
following suggestion: If the sum of the norms of the $m$ first
terms of the wavefunction fulfills the equation $\sum_{n=0}^{m}
N^{(n)}_{{\bf k}L} \cong 1$, then all terms with $n>m$ in the
Reiter's wavefunction can be neglected.

\begin{figure}[t!]
{\includegraphics[width=\textwidth]{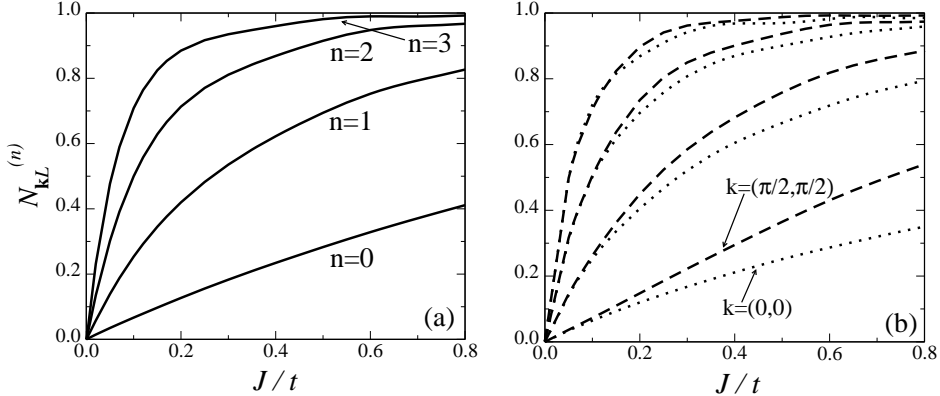}} 
\caption{The norm $N^{(n)}_{{\bf k}L}$ as a function of $J$ for
$n=0,1,2,3$, from bottom to top respectively, as obtained
for the orbital model (a) without and (b) with three-site 
terms Eq. (\ref{H3s_pol}) included in the Hamiltonian.
The results on panel (a) do not depend on ${\bf k}$ whereas dotted (dashed) lines 
on panel (b) show results for ${\bf k}=(0,0)$ 
[${\bf k}= (\pi / 2 , \pi / 2)$], respectively. All of the results do 
not depend on the sublattice index $L$.} \label{fig}
\end{figure}

In order to deduce what is the value of $m$ for different values
of the orbiton energy $J$ we calculated numerically $N^{(n)}_{{\bf
k}L}$ for $n=0,1,2,3$ as a function of $J$ on a mesh of $16 \times
16$ $k$ points \footnote{Obviously, in this procedure the QP
spectral weight $\Big|a^{(0)}_L\Big|^2$ is calculated directly from
the Green's function \cite{Ram94} and 
not from the normalization factor to the wavefunction.}, 
cf. Fig. \ref{fig}. The obtained results do not
depend on the sublattice index $L$. Fig. \ref{fig}(a) shows the
results for the orbital model Eq. (\ref{Ht_pol})--(\ref{HJ_pol})
whereas Fig. \ref{fig}(b) shows the results for the orbital model
Eq. (\ref{Ht_pol}-\ref{H3s_pol}). We conclude that: ($i$) for both
cases for $J\gtrsim 0.4 t$ the norm of the wavefunction is close
to $1$ already for $m=3$ and hence there are just up to three
orbitons participating in the formation of a polaron, ($ii$) it
seems that {\em only few more} orbitons are excited for smaller
$J$, and ($iii$) for the case {\em with} three-site terms
$N^{(n)}_{{\bf k}L}$ depends slightly on the momentum ${\bf k}$
--- it follows the dependence of the QP spectral weight $
N^{(0)}_{{\bf k}L} \equiv \Big| a^{(0)}_L({\bf k}) \Big|^2$ on ${\bf
k}$, with a maximum at ${\bf k} = (\pi / 2, \pi / 2)$,
i.e. at the minimum of $\varepsilon ({\bf k})$, cf. Eq.
(\ref{eq:disp}), and ($iv$) the dependence of $N^{(n)}_{{\bf k}L}$
on ${\bf k}$ found {\em only when} the three-site terms are
included demonstrates that the three-site terms are indeed
responsible for the formation of mobile polarons.

{\em Summary---} As a summary, it is instructive to compare the
results of the realistic orbital $t$--$J$ model (i.e. {\em with}
three-site terms included) with those obtained for the spin Ising
model and for the spin SU(2) symmetric model in Refs. \cite{Ram94}-\cite{Ram98}.
On one hand, for all $n$ the dependence of the norm $N^{(n)}_{{\bf
k}L}$ on ${\bf k}$ resembles to some extent the spin SU(2) model.
On the other hand, a closer look reveals that even for ${\bf k}=
(0, 0)$ the dependence of $N^{(n)}_{{\bf k}L}$ on $J$ is a concave
function for all $n$ just as for the spin Ising case and unlike in
the $SU(2)$ case where it can be a convex function of $J$ for some
$n$ \cite{Ram98}. Hence, the detailed study of the Reiter's
vavefunction of the orbital polaron confirms that the hole doped
into the $t_{2g}$ AO forms a {\em mobile} polaron just like the
mobile spin polaron in the AF plane of high-$T_c$ cuprates
\cite{Mar91} but its properties are truly distinct, in agreement
with discussion in Ref. \cite{Dag08}.

\section*{Acknowledgments}

This work was supported by the Foundation for Polish Science
(FNP), the~Polish Ministry of Science and Education under Project
No.~N202 068 32/1481, and the NSF under grant DMR-0706020.

\end{document}